# THE PHYSICS AND IDENTITY OF DARK ENERGY

Tom Gehrels

Space Sciences Building, University of Arizona, Tucson, AZ 85721-0092

ABSTRACT

This paper may solve the elusive dark-energy problem because our universe is not alone, and the multiverse is a powerful part of the cosmos. The decay of our aging universe is reviewed first. The accelerated expansion takes the decay debris into the inter-universal medium (IUM) of the multiverse, which conserves it on a $10^{30}$-years time scale. A prominent component of the debris and of the IUM is the enormous number of old cold photons from the decaying universes among the $10^{19}$ in our Local Group of universes.

The numerous old photons made the accretion possible of the large mass for our proto-universe, equivalent to $10^{21}$ solar masses, without black-hole collapse. This was accomplished by the energy-seeking property of the old debris material, and by the photon's *acceleration* property that counteracts some of the gravity.

When a central volume of our proto-universe reached $10^{18}$ kg m$^{-3}$ (proton density), and $10^{13}$ K, the old photons and protons became fully re-energized. That was a small central volume, apparently ~4.6 % of the total. Outside of that volume, the large numbers of remaining old photons continued their acceleration and the expansion of our universe.

The above accretion and expansion are described a second time with what we know about dark energy, particularly its *acceleration* of the expansion at age 5 x $10^9$ of our universe. Identical results are obtained; in fact, the two descriptions are complementary, and the conclusion is thereby made that dark energy is the acceleration energy of old photons.

There are two confirmations of the model. First, Karl Schwarzschild's limit allows this possibility (but not the Big Bang), and even confirms the mass of universes of our universe. Second, the old photons from decaying universes would have choked the system if they would *not* have been used; we would not have been here.

The model is supported by 30 observations, and by 19 considerations for future work, and that is a precious situation in present-day cosmology.

*Subject headings:* multiverse, universe, photons, dark energy, expansion.

## 1. INTRODUCTION

This arXiv.1101.0161 paper is a sequel to five papers since 2007 that are readily perused by selecting their http references in www.lpl.arizona.edu/faculty/gehrels2.html. A popular book has the assembly of all work on this model to date (Gehrels 2011a). The model is based on several observations, and its basic equation had been discovered by Subrahmanyan Chandrasekhar (1910-1995). I was in his class on stellar evolution in the 1950s and remembered how he stressed the possibility that the equation might show deeper relations between atomic theory and cosmogony; he published the equation at least four times (Chandrasekhar 1937, 1951, 1983, 1989). However, the time was not ripe to do anything with the equation until there would be consistent observations and understanding of our universe (Hinshaw 2010). An example is the discovery of the *acceleration of the expansion* (Riess et al. 1998; Perlmutter et al. 1999). The expected *de*celeration would have prevented this model, while *ac*celerated expansion brings the debris of our aging and decay into the inter-universal medium as material for new universes on a time-scale of $10^{30}$ years.

This paper explores if a model based on such a powerful equation can tell us the physical nature of the major component of our universe, dark energy. And indeed, a solution to the dark-energy problem is readily found; it is rather model independent - valid for any multiverse, not just the Chandra Multiverse - as follows.

Everything ages and decays - protons have a half-life of at least $10^{35}$ years - and the debris floats on the accelerated expansion of intergalactic space into the inter-universal medium (IUM), in which our universe is imbedded. When a new universe is accreted from an IUM cloud to make a proto-universe, the photons and protons eventually reach $10^{13}$ K and proton density of $10^{18}$ kg m$^{-3}$; these are parameters from atomic theory for them to be completely re-energized. That is the beginning of a new universe like ours, at $10^{-6}$ s on the old clock ($10^{37}$ Planck times beyond a Big Bang); as for what happened earlier, the subatomic physics of $<10^{-6}$ s are the domain of the multiverse.

This scenario is tried a second time with what we know about dark energy. The result of the events is the same, more complete in fact. *Dark* energy is therefore another name for the acceleration energy by *old* photons; their lack of observability is in Sec. 3.1.

Section 2 first shows the powerful physics for the model, and Sec. 3 the history of our universe; these sections are brief because their details are in the above five papers. Section 4 has the physics for the beginning of our universe, which leads to the discovery of the physics of dark energy in Sec. 5. Section 6 shows two independent confirmations. Section 7 summarizes 30 supporting observations and considerations, Sec. 8 lists future work, and Sec. 9 has a summary. This paper is followed by arXiv.1101.0567, on the physics and identity of dark matter.

## 2. THE PHYSICS OF THE MULTIVERSE

Chandrasekhar (1951)**,** in his course on the structure, the composition, and the source of radiation of the stars, derived an equation for masses in our quantized universe,

$$M(\alpha) = (hc/G)^{\alpha} H^{1-2\alpha}, \qquad (1)$$

in which h is the Planck constant, c the velocity of radiation, G Newton's gravitational constant, and H the mass of the proton; it is for positive exponents $\alpha$ identifying the objects shown in Table 1.1. From his derivation of Eq. (1), it appears to have quantum, relativity, gravity, and atomic physics unified in the cosmos, and that is awesome. Planck (1899) expressed such awe, and that was merely for his constant h, with,

> "... the possibility is given to establish units for length, mass, time and temperature, which, independent of special bodies or substances, keep their meaning for all times and for all cultures, including extraterrestrial and non-human ones, and which therefore can be called 'natural measurement units' "

The unit we are interested in here is for mass, $(hc/G)^{0.5}$, now called the Planck mass; he had obtained them all by dimensional analysis, while Chandra's derivation calibrated them. The second term of Eq. (1) serves merely so that any mass unit can be used, such as kg or solar masses. A "universal Planck mass" appeared when I simplified Eq. (1) by expressing all masses in terms of the *universal* unit of the proton mass, such that H = 1, and

$$M(\alpha) = (hc/G)^{\alpha}. \qquad (2)$$

Chandra had already made some comparison with observations, I pursued that, and also found that the objects in Table 1.1 are the only ones participating in M(α), with the possible exception for planetesimals at α = 1.00. The equations are however open to higher values of α; only the first quantization step in the multiverse is shown in the Table. Another way to show the quantization is in whole steps that emphasize the Planck mass,

$$M(N) = (hc/G)^{0.5N}. \qquad (3)$$

The Table has values computed with these three expressions, and the last column shows the result of comparisons with observations.

TABLE 1.1

MASSES DERIVED FROM THE ABOVE EQUATIONS

| α | N | Predicted Proton masses | Units shown | Type of Object | Observed α |
|---|---|---|---|---|---|
| 2.50 | 5 | $3.68 \times 10^{97}$ | $3.26 \times 10^{19}$ un | Local Group of Universes | - |
| 2.00 | 4 | $1.13 \times 10^{78}$ | $9.52 \times 10^{20}$ s. m. | Primordial Universe | 1.998-2.008 |
| 1.50 | 3 | $3.47 \times 10^{58}$ | 29.179 s. m. | O and B stars | 1.49-1.53 |
| 0.50 | 1 | $3.26 \times 10^{19}$ | $5.46 \times 10^{-8}$ kg | Planck mass | 0.50 |
| 0.00 | 0 | 1 | $1.67 \times 10^{-27}$ kg | Proton | 0.00 |

un = universes; s.m. = solar masses

## 3. THE HISTORY OF OUR UNIVERSE

### 3.1. *The Ending of our Universe*

The ending is considered first because the participating components are then easily identified for usage in the multiverse and proto-universe. Everything in a universe ages and decays; even the proton may have a limited half-life.

Old cold protons and other particles such as neutrons and electrons are part of our universe's decay debris, as are whole galaxies (each gravitationally holding its debris), clusters of galaxies, and whatever other debris such as old stars. Their ensemble is defined as *protons etc*. In addition, dark matter and dark energy are abundant in our universe.

Photons are especially important in this scenario; their properties are curiously unique, as if they evolved in the multiverse for the role we see them play (Lamb 1995). Foremost is their *acceleration*: with velocity always c, they impart momentum in collisions with objects they encounter; a solar sail is a special example, and we will discuss the interaction with protons later. The photons seem to vanish when and only when their energy has been transferred through acceleration to the charged particles or other objects. Apparently, they keep all but one of their properties at all temperatures; the exception being the radiation depending on temperature. The declining amount of radiation of the "old photons" is observed in terms of cooling in space on the expansion of our universe; COBE and other surveys observed it at 3000 K for age 380,000, and near 3 K at present. During their long stay in the multiverse, they must be near absolute zero.

### 3.2. *In the Chandra Multiverse*

If our universe decays into the inter-universal medium (IUM) - and had originated from it - the IUM will have our physics. The IUM has uniformity through mixing of debris from a large number of universes, which had all emerged from the IUM to begin with, such that they all have that h, c, G, H physics.

The fundamental characteristic of the multiverse is that it is an *evolutionary system* (Gehrels 2011b). Evolution considers slightly different characteristics of nature's components, occurring by chance, caused by the environment in *natural selection*. Will the species survive with them, or is the difference too great so as not to survive? One speaks then of *trial-and-error evolution*. A universe may have characteristics that deviate too much so that it will not survive, and it vanishes back into the IUM. Or, a universe may emerge with slightly different characteristics.

In this manner, the multiverse evolved the physics of the universes, like ours, which is so finely tuned that it can produce stars, planets, and people. That extreme fine-tuning has been a puzzle for decades for it could not have happened in a single and rapidly developing early Big-Bang universe. Fred Hoyle's name is associated with that, how he wondered, in books and lectures, about the extreme fine-tuning of atomic specifications in stars. Now we know however that the multiverse has many samples of universes, $10^{19}$ in the Local Group, that evolve for long times, estimated at $10^{30}$ years; they should then be highly successful in evolution, as Darwin's finches and people demonstrate.

### 3.3. *The Beginning of our Universe*

Clouds probably form in the IUM as in the interstellar medium (ISM). The supply from many aging and decaying universes into the IUM is on average uniform as well as continuous, even though the universes are born at randomly different times depending on when their cloud began to form. Nothing stands still in the cosmos - the clouds continue to grow by sweeping the material up during their motion through space. Eventually, self-gravitation will become active, speeding the accretion of the cloud by its increasing gravitational cross-section.

There is a classical problem that a cloud of mass equivalent to $10^{21}$ solar masses would be far too massive, it would get too hot and all characteristics would be melted away; it would probably collapse into a black hole. Now however, there are two causes why that problem may have been solved. First, *the acceleration pressure* of photons provides counter-action to gravity, as we shall see in detail in Sec. 4. Second, the IUM *composition* differs from that in the ISM, namely of the above decayed *energy seeking* debris, instead of the usual atomic and molecularly active ISM material.

The growing proto-universe therefore has a maximum temperature of $10^{13}$ K, which is apparently all right for the mass concentrations of galactic clusters in our debris to have survived in our universe (#14 in Sec. 7).

Imagine a gravitationally spherical cloud of about the size of the Venus orbit having uniformly mixed debris. Gravitational energy of compaction was used to re-energize the old photons and old atomic components into regular photons, protons, neutrons, electrons, etc. When in accretion the *central* volume obtained the density of $10^{18}$ kg m$^{-3}$ - which is the density at which photons and protons formed in standard modeling - the old photons and protons were fully re-energized. There is the peculiar fact that only 4.6% of the mass of our universe is baryonic, our visible matter. Why it is so small has been a puzzle, but now it is seen in the simple spherical geometry of a gravitational globe. With highest density at the center, a maximum-density central mass will be a small fraction of the whole; it apparently was only 4.6 %.

Outside of the 4.6% volume, the re-energizing and re-constitution did not come to completion because of insufficient density. Old photons and old protons etc. remained at their high percentages and they continued their acceleration interaction in multiple scattering, sustaining expansion.

Back to the central region, it had also its old *protons* re-energized; $10^{18}$ kg m$^{-3}$ actually is proton density, at $10^{13}$ K, while t ~ $10^{-6}$ s is the epoch of their formation in standard modeling. The new photons interacted with the new protons in multiple scattering, also by electrons and neutrons. The epoch of t ~ $10^{-6}$ s is the beginning of our universe, ~$10^{37}$ Planck times away from a Big Bang model - our universe began at the photon and proton level. Because of the difference of starting our universe, by as many as $10^{37}$ Planck times, this model need not and does not consider Strings, Inflation, or early Big Bang theories. The only WMAP observation supporting Inflation theory may just as well support the Chandra Multiverse, for the sudden appearance of a Venus-orbit sized baryonic object at t ~ $10^{-6}$ s.

The physics of before t ~ $10^{-6}$ s (that already had been evolved in the multiverse, over long times) was maintained in the multiverse at this epoch, as it always is; this is of course the h, c, G, H physics of the Chandra Multiverse Model (Sec. 3.2).

After t ~ $10^{-6}$ s, the multiple scattering and expansion continued throughout the universe as well as in the central volume. That lasted until age 380,000, when at space density of ~$10^{-19}$ kg m$^{-3}$ the electrons, protons, and neutrons combined to make hydrogen and helium atoms, which have wide internal spacing for the photons to escape through.

## 4. THE CAUSE OF EXPANSION OF OUR UNIVERSE

In the previous section, the discussion before the epoch of t ~ $10^{-6}$ s had *accretion* in our proto-universe, and *expansion* afterwards. What caused the reversal? "What caused the expansion?" is the classic question.

Already during the formation of our proto-universe, there was an increasing effect of an outward force on the protons etc. The force was due to the *acceleration* by the enormous number of old photons from the decaying universes; *i.e*, the photons exerted pressure outward, a *negative* effect with respect to a positive direction for the inward direction of gravity.

The effect was increasing because the scooping up of material from the IUM depended on the cross section of the cloud. However, its pressure always *lagged behind* the gravitational force (as the density increased from ~$10^{-28}$ kg m$^{-3}$ in the IUM to the proton density of $10^{18}$ kg m$^{-3}$). This is because it brought the appropriate temperatures from 0 to $10^{13}$ K all right, but always lagging behind, taking time because the cold material had to be heated. The gravity thereby prevailed and the accretion continued, but the dominance of the gravity gradually slowed down.

It did come to a remarkable stop by a reversal from accretion into expansion when at the center of the cloud the density reached $10^{18}$ kg m$^{-3}$ at which in the standard model the births of photons and protons occur (Sec. 3.3). There may in fact have been a short interval in which the old photons were re-energized while for the protons etc, it may have taken a little longer because some of them may have had to be re-constituted as well. If so, there was the Photon Burst of the $10^{13}$-K fireball energetic enough to be observed by WMAP as a radiation signature with a *wider* curvature than that of the 3-K radiation; that is of a burst of the radiation well *before* age 380,000 (Hinshaw 2010).

Both these bursts would have pushed the accreting material away from the proto-universe; the blasts had a pressure blowing the IUM material away from the completed universe. It will be exciting to do the detailed computer modeling; in outline, it has already been done in the Schwarzschild part of Sec. 6, showing the cause for all universes to have our universe's mass M.

Continuing with the beginning-of-our-universe scenario, after the t ~ $10^{-6}$ s events the acceleration by the photons inside the 4.6-% volume sustained the *expansion*, as did the old photons outside of that volume. However, now the appropriate temperature from $10^{13}$ K to 3,000 K was always *lagging behind* the decrease of the density from $10^{18}$ to $10^{-19}$ kg m$^{-3}$, because the cooling took time. The expansion was therefore leading as the pressure declined.

After t ~ $10^{-6}$ s (equal to t = 0 on the Chandra-Multiverse clock), our understanding of the physical evolution appears to be on the track of the standard models for our universe and for atomic physics. Our universe had begun with photons and protons, to continue on the path described by the standard models.

## 5. THE DISCOVERY

This section is to describe again the events leading up to age 380,000, but now not in terms of old photons but with what is known about dark energy. The following percentages are in terms of mass (Hinshaw 2010):
1. At the present time, the baryons amount to only 4.6%. Neutrinos have less than 1%, while 23% is not-observable dark matter. The dominant 72% is in some form of dark energy.
2. When the universe had age 380,000, it amounted to 12% atoms, 15% regular photons, and 10% neutrinos. Not observable but otherwise derived to be present was 63% dark matter. There was little or no dark energy.

While dark energy is presently dominant at the above 72 %, our decaying universe and the inter-universal medium would have kept at least that percentage, and so would the accreting proto-universe. In contrast, the above Point 2) shows that at age 380,000 there was little or no dark energy. Apparently, it had been used up.

That difference of the two dark-energy observations in Points 1) and 2) can therefore be used to answer the question again, "What caused the expansion of our universe?" *Causing* the expansion has the same physical action as *accelerating* the expansion at age 5 x $10^9$, which is in the literature as due to dark energy (Riess et al. 1998). The conclusion is therefore that the original expansion is caused by dark energy as well as is the later acceleration.

Independent of that conclusion, we have the same course of events and results with dark energy as with the above old photons. "Dark energy" is therefore a name for what is the physical action of acceleration by old photons. The discovery is that *dark energy is the energy of old photons*.

## 6. TWO CONFIRMATIONS

The above two sections are ideal for computer modeling, as was done for the Chandrasekhar limit to correct it to 1.4 solar masses; presently, the entire model needs to be verified.

The dark-energy conclusion has a simple but firm confirmation. The evolution in the multiverse could not have left unused the major part of what all decaying universes contribute to the IUM, namely their old photons (or, in the second interpretation, their 72% dark energy). *If* these would have been left unused, the ever-increasing number of old-photon debris (*i.e.* dark energy) would have choked the multiverse.

A confirmation comes from the warning initiated by Schwarzschild (1916). It is seen simply as a limiting radius of the proto-universe, $R_S$, for light to escape, at which the local gravity is equal to or less than the energy of the light,

$$R_S = 2GM\,c^{-2}. \tag{4.1}$$

Table 6.1 shows the comparison of $R_S$ with radius $R = (3M/4\pi\sigma)^{1/3}$ for a sphere with uniform density $\sigma$. $M = 1.13 \times 10^{78}$ proton masses, and it is noted that the Table thereby is a confirmation of the entire model, specifically that all universes of the multiverse have the same mass, M.

TABLE 6.1

RADII AND SCHWARZSCHILD RADII

| $R/R_S$ | t | $\sigma$ | R(ly) |
|---|---|---|---|
| 1 | 380,000 y | $10^{-23}$ | $10^8$ |
| $10^{-14}$ | $10^{-6}$ s | $10^{18}$ | $10^{-5}$ |
| $10^{-40}$ | 0 | $10^{96}$ | $10^{-31}$ |

The first line applies when the universe's radiation is known from spacecraft observations to escape at t ~ 380,000 years, $R/R_S = 1$. The modeling does not seem precise because standard theories predict the density to be ~$10^{-19}$ kg m$^{-3}$ at that time, not $10^{-23}$. However, the precision of these predictions is low, and their effect is small because if $10^{-19}$ would be used in the calibration of $R_S$, the ratios $R/R_S$ would still be $10^{-14}$, and the other one $10^{-38}$ instead of $10^{-40}$.

The second line is for the above t ~ $10^{-6}$ s, using proton density of $10^{18}$ kg m$^{-3}$ to derive R. Because $R/R_S = 10^{-14}$ is so strongly negative, the warning indicates a black hole.

However, our universe is not a black hole. Furthermore, for photons to escape at age 380,000, they must have been *generated much earlier*; the 380,000 event must have been long in preparation. In the case of the sun, escape takes a million years for a photon generated at its center. That is physically a different situation, but it may give some indication that the spell of 380,000 years being shorter than a million, for a body much more massive than the sun, follows from the fact that its medium was *expanding*. Indeed, the scattering density decreased rapidly with time, to as low as the above $10^{-19}$ kg m$^{-3}$ at age 380,000.

In the third line, the Planck density of $10^{96}$ kg m$^{-3}$ is used (as if our universe were ever at t = 0 on the Big-Bang clock), for obtaining R and thereby $R/R_S$. With $R/R_S = 10^{-40}$, the warning calls resoundingly for a black hole. However, our universe is not a black hole, while this time neither photons nor dark energy are available to save the explanation.

## 7. SUPPORTING OBSERVATIONS AND CONSIDERATIONS

Here is the list of observations used thus far in the modeling. With an asterisk * are indicated the ones that have not as yet been verified because that can be done only by experts familiar with the data processing of the WMAP observations. On the other hand, with an elevated $^{BB}$ are the three often-quoted proofs of the Big Bang theory, which support this model just as well because they are for epochs later than t ~ $10^{-6}$ s.

1. The understanding of the inter-universal medium (IUM) is based on extensive observations of the interstellar medium.

2. The universes photons decay into "old cold photons", which are observed by WMAP from 3,000 K at age 380,000, down to 3 K now.

3. The discovery of the acceleration of expansion at age ~5 x $10^9$ is essential for this model; the model could not exist if *de*celeration had been found, while now the expansion appears to move debris outward into the IUM.

4. The Chandra equation of Eqs. (1) and (2) has at its foundation the proton and Planck masses, which are observed in terrestrial laboratories.

5. The quantization factor of 3.3 x $10^{19}$ is observed between the masses of our universe and its O-stars, and between Planck and proton masses.

6. The computed mass of our universe yields the equivalent spherical radius of the proton, 8.2 x $10^{-16}$ m, on the mean of three sets of observations (this brings a special consideration regarding the spread of the precise observations and the shape of the proton).

7. The mass of O-type stars is observed near the computed 29.2 solar masses.

8. The mass of our universe is close to the critical mass, as it must be in order to have survived in the multiverse.

9. That there has to be an evolutionary multiverse rather than a single universe is supported by inorganic evolution and its many observations (Gehrels 2007).

10. Section 3.1 has about photons, "their properties are curiously unique as if they evolved in the multiverse for the role we see them play". This is, of course, the case, as it is for every property of all surviving components of the multiverse.

11. The Supply Problem is solved: where did our universe's observed energy equivalent to $10^{21}$ solar masses come from?

12. The origin of our finely tuned physics is solved, because the multiverse has the many universes and long times required for finely tuned evolution.

13. The First Uniformity Problem: uniformity is observed until third-decimal precision in the 3-K background observations by COBE; the temperature of 1.728 K is observed in all directions. Here it is understood from the inputs into the IUM from many decaying universes.

14. The Second-Order Non-Uniformity Problem: how could the fifth decimal of the background show appreciable non-uniformity with variations on a scale of galactic clustering, observed in the WMAP maps (Spergel et al. 2007)? This is understood because the galactic-clustering debris was input into the IUM to begin with, and the $10^{13}$ K temperature apparently was not too high to homogenize that mass distribution.

15. It did not make sense in the case of the old BB universe to ask where all its radiations went; but a surrounding multiverse conserves them.

16. In the multiverse, the old photons cannot stay and pile up - they must be used. And they are used, for there is great need for them, to make the accretion of our Proto-Universe possible, that is of the equivalent of $10^{21}$ solar masses, without collapse into a black hole.

17. The inertness and energy-seeking properties of the old universes' debris also played a role to make this possible, keeping the temperature down to $10^{13}$ K.

18. The simple consideration of a small central volume at highest density in a globe shows why the percentage of baryons is small, only 4.6%.

19. The timing of $t \sim 10^{-6}$ s for the beginning of our baryon universe is from the standard model of atomic physics as the time of formation of photons and protons, at proton density $10^{18}$ kg m$^{-3}$.

20*. The excitement for Inflation Theory came when WMAP confirmed it in part (but not for an infinite inflation), while for the present model that same confirmation might fit the sudden appearance of a 0.6-AU baryonic object, of protons and neutrons at $t \sim 10^{-6}$ s (old clock), t = 0 now.

21*. The Photon Burst, a huge fireball of $10^{13}$ K, may be confirmed by the WMAP observations of a radiation signature with a wider curvature than that of the 3-K radiation.

22. Following the Photon Burst, if there was one, the re-energized photons were being scattered by re-energized and perhaps re-constituted protons, electrons, and neutrons, so as not to get out until age 380,000 years – the appreciation of this time period is based on solar observations.

23. The observation that at age 380,000 the universe was expanding is basic to understanding the proto-universe.

24$^{BB}$. The expansion of our universe used as a proof for the Big Bang, applies to this model for $t > 10^{-6}$ s just as well.

25$^{BB}$. The same applies to the 3-K cosmic background.

26$^{BB}$. The same applies to the successful prediction of the hydrogen-to-helium ratio.

27. The Schwarzschild test confirms this model; the test's basic observation is that our universe is not a black hole.

28. The dark energy appears from WMAP observations to have been used up by age 380,000.

29. Dark energy is observed as old photons.

30. Dark matter is observed as old cold protons and other particles such as neutrons and electrons as part of our universe's decay debris, as are whole galaxies (each gravitationally holding its debris), clusters of galaxies, and whatever other debris such as old stars; the ensemble is referred to as "protons etc."

## 8. RESULTS AND FUTURE WORK

Whenever I learned of a new observation, it always fit into the model, and this seems an indication of its truth, as does its aesthetics (Chandrasekhar 1987). With so many observations in support – a rare case in this discipline – the model should be close to the cosmos' truth, and this is seen in the following 19 examples.

The theory is however only approximate, and the points therefore also indicate future work. For example, a numerical modeling of the above scenarios would be of the greatest interest, even though it is already noted with Table 6.1 that it is a confirmation of the entire model, and shows the same mass for all universes in the multiverse.

• There was strict application of an equation, without anthropic assumptions. The results of the model indicate that observations do not affect quantum theory for the macro-cosmos.
• The quantization of our universe is demonstrated, and extended into the multiverse, as it is based on proton and Planck masses.
• The $M(\alpha)$ equation is connected to the Planck domain and is recognized as a *universal* Planck mass.
• The $M(\alpha)$ equation uses quantum, relativity, gravity, and atomic physics together in a unified manner, as its derivation shows (Chandrasekhar 1951).
• This is a specific multiverse model, namely a hierarchy of universes that have the same mass and physics, based on the $M(\alpha)$ equation.
• The model provides a history for the beginning and ending of our universe.
• It uniquely and specifically overcomes the Schwarzschild warnings regarding the reality of the early phases of Big-Bang models.
••• The Horizon Problem, Flatness Problem, and Supply Problem are presently solved by an *ad hoc* inflation (Padmanabhan 2002), while in the specific multiverse they are *naturally* solved by the accretion of debris from many universes for our proto-universe.
• The deviations from uniformity at higher than second-decimal precision of the Cosmic Microwave Background follow naturally from the conservation of clustering of galaxy masses throughout the procedures of decay, accretion, and universe formation.
• The model brings up the origin of our physics in the multiverse.
• The cosmological foundation of our world and its physics is in trial-and-error evolution within its hierarchy of universes.
• The multiverse evolution found the cosmos' h, c, G, H physics.
• The modeling found a logical origin for our universe, based on observations, that does not invoke a mysterious point in time called the Big Bang.
• The classical wondering why the baryons have only 4.6% of the composition follows readily from spherical geometry of our proto-universe.
• The treatment of this paper indicates that the Planck constant is h, not $\hbar = h/2\pi$, or the O stars and universe masses would be unacceptable; a better name for $\hbar$ may be the Dirac constant.
• It also implies that the cosmological constants h, c, G, and H are constant with time and location.

• Fred Hoyle's fine-tuning of the nuclear transitions within stars is explained by the trial-and-error evolution within the multiverse, over $10^{30}$ years and $10^{19}$ sample universes, producing finely tuned universes to begin with.
• Because the universes begin at $\sim 10^{-6}$ secs instead of t = 0, all problems of the earliest Big Bang are avoided.

## 9. CONCLUSIONS

A universe generates many photons and old cold photons, and WMAP observes the latter at 3000 and 3 K. The accelerated expansion brings the debris of our decaying universe into the Chandra Multiverse, where they are conserved cold for long times.

Eventually, there is formation for the cloud that would become our proto-universe; it has the old photons still at high percentage of the total composition, ~72%. The acceleration property of the photons diminishes the effect of gravity through multiple scattering on the other components; the radiation property increases because of the rising temperature. When proton density is reached in a central volume, there may be a Photon Burst explosion; the old protons are re-configured and they increase the multiple scattering of the photons. That central volume apparently has 6.4% of the volume, *i.e.,* our baryon percentage; the explosion may be the one observed by spacecraft as a radiation signature with a wider curvature than that of the 3-K radiation. Outside of the "6.4% volume", the old photons and protons, etc. continue their multiple scattering, which sustains the expansion.

The scenario is played twice, with photons and with what is known about dark energy, and the results are identical and even complementing. Dark energy is therefore a name for the acceleration energy of photons. The statistical confirmation of this explanation is that the photons from all old universes cannot have stayed in the multiverse, or our multiverse would not be as we know it today. We would not have been here; we ourselves are the supporting observation for the conclusions.